\documentclass[11pt,letterpaper]{amsart}

\usepackage[left=1in,top=1in,right=1in,bottom=1in]{geometry}
\usepackage[foot]{amsaddr}
\usepackage{amsmath}
\usepackage{amsthm}
\usepackage{amssymb}
\usepackage{url}
\usepackage{hyperref}
\usepackage{mathtools}

\pdfoutput=1

\newtheorem{Thm}{Theorem}
\newtheorem{Lem}[Thm]{Lemma}

\theoremstyle{remark}

\theoremstyle{definition}

\theoremstyle{definition}
\newtheorem{Hyp}[Thm]{Hypothesis}

\newenvironment{Proof}{\begin{proof}}{\end{proof}}

\newcommand{\bra}{\ensuremath{\langle}}
\newcommand{\ket}{\ensuremath{\rangle}}
\newcommand{\F}{\ensuremath{\mathbb{F}}}
\newcommand{\Z}{\ensuremath{\mathbb{Z}}}
\newcommand{\Nat}{\ensuremath{\mathbb{N}}}
\newcommand{\SAT}{\ensuremath{\mathrm{SAT}}}
\newcommand{\poly}{\ensuremath{\operatorname{poly}}}
\newcommand{\polylog}{\ensuremath{\operatorname{polylog}}}

\begin{document}

%%%%%%%%%%%%%%%%%%%%%%%%%%%%%%%%%%%%%%%%%%%%%%%%%% Front-matter information %%%

\title[]{The Fine-Grained Complexity of Computing\\{}
         the Tutte Polynomial of a Linear Matroid}

\author{Andreas Bj\"orklund}
\address{This work was carried out while AB was employed as a researcher at Lund University, Department of Computer Science,
and the major part of the writeup was carried out while AB was employed as a researcher at Ericsson Research.}
\email{andreas.bjorklund@yahoo.se}

\author{Petteri Kaski}
\address{Aalto University, Department of Computer Science}
\email{petteri.kaski@aalto.fi}

\begin{abstract}
We show that computing the Tutte polynomial of a linear matroid of 
dimension~$k$ on $k^{O(1)}$ points over a field of $k^{O(1)}$ elements 
requires $k^{\Omega(k)}$ time unless the \#ETH---a counting extension of 
the Exponential Time Hypothesis of Impagliazzo and Paturi~[CCC~1999] 
due to Dell~{\em et~al.}~[ACM~TALG~2014]---is false. This holds also for 
linear matroids that admit a representation where every point is associated 
to a vector with at most two nonzero coordinates. Moreover, we also show that 
the same is true for computing the Tutte polynomial of a {\em binary} matroid 
of dimension $k$ on $k^{O(1)}$ points with at most three nonzero coordinates 
in each point's vector.
These two results stand in sharp contrast to computing the Tutte polynomial 
of a $k$-vertex graph (that is, the Tutte polynomial of a {\em graphic} 
matroid of dimension $k$---which is representable in dimension $k$ over 
the binary field so that every vector has exactly two nonzero coordinates), 
which is known to be computable in 
$2^k k^{O(1)}$ time~[Bj\"orklund~{\em et al.}, FOCS~2008]. 
Our lower-bound proofs proceed in three steps:
\medskip
\begin{enumerate}
\item
a classic connection due to Crapo and Rota [1970] between the number of 
tuples of codewords of full support and the Tutte polynomial of the 
matroid associated with the code;
\item
an earlier-established \#ETH-hardness of counting the solutions to a 
bipartite $(d,2)$-\textsc{CSP} on $n$ vertices in $d^{o(n)}$ time; and
\item
new embeddings of such \textsc{CSP} instances as questions about codewords 
of full support in a linear code.
\end{enumerate}
\medskip
Geometrically, our hardness results also establish that it is \#ETH-hard to 
compute the volume of proper hyperplane chambers in time $k^{o(k)}$ for 
a given arrangement of hyperplanes through the origin of a finite 
$k$-dimensional vector space over a $k^{O(1)}$-element field.
We complement these lower bounds with two algorithm designs to form essentially
a complexity dichotomy under \#ETH. The first design computes the Tutte 
polynomial of a linear matroid of dimension~$k$ on $k^{O(1)}$ points in 
$k^{O(k)}$ arithmetic operations in the base field. The second design 
generalizes the Bj\"orklund~{\em et al.} algorithm from the graphic case and 
runs in $q^{k+1}k^{O(1)}$ time for linear matroids of dimension $k$ 
defined over the $q$-element field by $k^{O(1)}$ points with at most 
two nonzero 
coordinates each.
\end{abstract}

\maketitle

\clearpage

%%%%%%%%%%%%%%%%%%%%%%%%%%%%%%%%%%%%%%%%%%%%%%%%%%%%%%%%%%%%% Document body %%%

\section{Introduction}

\subsection{Matroids and the Tutte polynomial}
A {\em matroid} is a tuple $(E,\mathcal I)$, where $E$ is a finite set 
of \emph{points}, and $\mathcal I$ is a nonempty set of subsets 
of $E$ called the \emph{independent sets} of the matroid with the following 
two properties:
\begin{enumerate}
\item 
every subset of an independent set is an independent set; and 
\item 
for any two independent sets $A$ and $B$ with $|A|>|B|$, 
there exists an $e\in A\setminus B$ such that $B\cup\{e\}$ 
is an independent set.
\end{enumerate}
Matroids generalize fundamental combinatorial and algebraic notions such 
as graphs and linear independence in vector spaces; for an introduction, 
cf.~Welsh~\cite{Welsh1976} and Oxley~\cite{Oxley2011}.

A matroid is {\em linearly representable} (briefly, {\em linear}) over a 
field $\F$ if it can be described by a $k\times m$ matrix 
$M\in \F^{k\times m}$ of rank $k$, where the number of rows $k$ is 
the \emph{dimension} of the matroid, and the $m$ columns are indexed by 
the points $E$ of the matroid with $|E|=m$.
For any subset $S\subseteq E$ of the columns, let us write $M[S]$ to denote 
the $k\times |S|$ matrix obtained by restricting $M$ to the columns 
indexed by $S$. We write $\rho(S)$ for the rank of $M[S]$ over $\F$. 
The independent sets of a linear matroid are the sets $S$ for which 
$\rho(S)=|S|$; that is, the subsets of linearly independent vectors.

The {\em Tutte polynomial} of a linear matroid $M$ is the integer-coefficient
polynomial in two indeterminates $x$ any $y$ defined by
\begin{equation}
\label{eq:tutte}
T_M(x,y)=\sum_{S\subseteq E} (x-1)^{k-\rho(S)}(y-1)^{|S|-\rho(S)}\,.
\end{equation}
This generalisation of the Tutte polynomial from graphs to matroids was
first published by Crapo~\cite{Crapo1969}, although it already appears
in Tutte's thesis; Farr~\cite{Farr2007} gives an historical account of 
the Tutte polynomial and its generalizations. 
Brylawski~\cite{Brylawski1972}---foreshadowed by 
Tutte~\cite{Tutte1947,Tutte1976}---showed that the Tutte
polynomial is a universal invariant for deletion--contraction recurrences, 
and thus captures a wealth of combinatorial counting invariants ranging
from the chromatic polynomial of a graph to partition functions in
statistical physics to weight enumerators in coding theory;
cf.~Biggs~\cite{Biggs1993}, Godsil and Royle~\cite{GodsilR2001}, 
Vertigan~\cite{Vertigan1998}, and 
Welsh~\cite{Welsh1993} for a detailed account. 
Among these connections, the most relevant to our present work is the 
connection of the Tutte polynomial to full-support tuples of codewords of 
in linear code, cf.~Sect.~\ref{sect:techniques} for a discussion.

In 2008, Bj\"orklund {\em et al.}~\cite{BjorklundHKK2008} showed that if 
the matroid is \emph{graphic}; that is, when the matrix $M$ is an incidence 
matrix of an undirected graph over the binary field, then the Tutte 
polynomial can be computed in time $2^k\poly(k,m)$. Due to the universality
of the Tutte polynomial, it would be highly serendipitous to obtain 
a similar running time for a larger class of matroids.

\subsection{Our results---fine-grained dichotomy for the Tutte polynomial}

In this paper, we prove that such a running time for two natural ways 
of extending the graphic case to a larger class of linear matroids would have 
unexpected consequences in the fine-grained complexity of counting. 
Namely, we relate 
the complexity of computing Tutte polynomials of linear matroids to 
the {\em Counting Exponential Time Hypothesis} (\#ETH)%
---cf.~Sect.~\ref{sect:num-eth} for a precise statement---of 
Dell~{\em et al.}~\cite{DellHMTW2014}, which relaxes the 
Exponential Time Hypothesis (ETH) 
of Impagliazzo and Paturi~\cite{ImpagliazzoP1999}. 

Our first main theorem shows that under \#ETH one cannot extend the graphic 
case---that is, the binary field with at most two nonzero entries in every 
column of $M$---to moderately large field sizes without super-exponential 
scalability in $k$.

\begin{Thm}[Hardness of Tutte polynomial of a linear matroid under \#ETH]
\label{thm:main-lb}
Assuming \#ETH, there is no deterministic algorithm that computes 
in $k^{o(k)}$ time the Tutte polynomial of a given linear matroid 
$M$ of dimension $k$ with $k^{O(1)}$ points over a field of size $k^{O(1)}$. 
Moreover, this holds even when every column of $M$ has at most 
two nonzero entries.
\end{Thm}

Our second main theorem shows that under \#ETH one cannot extend the 
graphic case to more general matrices even over the binary field (or any
other fixed finite field) without super-exponential scalability in $k$.

\begin{Thm}[Hardness over any fixed finite field under \#ETH]
\label{thm:main-lb-fixed}
Assuming \#ETH, there is no deterministic algorithm that computes in
$k^{o(k)}$ time the Tutte polynomial of a given linear matroid $M$ of
dimension $k$ with $k^{O(1)}$ points over any fixed finite field.
Moreover, this holds even when every column of the matrix $M$ has at
most three nonzero entries.
\end{Thm}

We complement these hardness results to essentially a complexity dichotomy
under \#ETH with two algorithm designs. The first design is a deterministic 
algorithm for linear matroids, but with super-exponential scalability in 
the dimension $k$.

\begin{Thm}[An algorithm for general linear matroids]
\label{thm:main-ub}
There is a deterministic algorithm that computes the Tutte polynomial
of a given linear matroid $M$ of dimension $k$ with $k^{O(1)}$ points over
a~$q$-element field in time $k^{O(k)}\polylog q$ and $k^{O(1)}\polylog q$ space.
\end{Thm}

The second design is a deterministic algorithm for the case when 
each point has at most two nonzero coordinates, with field-size-dependent 
exponential scalability in $k$. In particular, this algorithm 
generalizes the $2^kk^{O(1)}$-time algorithm of Bj\"orklund \emph{et al.}~\cite{BjorklundHKK2008} from the 
graphic case to arbitrary finite fields.

\begin{Thm}[An algorithm for weight at most two]
\label{thm:main-ub-wt2}
There is a deterministic algorithm that computes the Tutte polynomial
of a given linear matroid $M$ of dimension $k$ with $k^{O(1)}$ points, each 
having at most two nonzero coordinates over a $q$-element field, in 
$q^{k+1}k^{O(1)}$ time and space.
\end{Thm}

Previously, the hardness of the Tutte polynomial has been studied restricted
to the graphic case from a number of angles, including the \#P-hardness 
results of Jaeger, Vertigan and Welsh~\cite{JaegerVW1990} 
(see also Welsh~\cite{Welsh1993}), the counting inapproximability
results of Goldberg and Jerrum~\cite{GoldbergJ2008}, the fine-grained
hardness results of Dell {\em et al.}~\cite{DellHMTW2014} under \#ETH, 
as well as the fine-grained dichotomy results of Brand, Dell, and 
Roth~\cite{BrandDR2019}. 
\subsection{Key techniques---linear codes and sparse algebraic constraint satisfaction}

\label{sect:techniques}

Let us now give a high-level discussion of the key techniques employed.
We proceed to prove Theorems~\ref{thm:main-lb} and~\ref{thm:main-lb-fixed} by
utilizing known connections between linear codes and the Tutte polynomial.
Towards this end, let us recall some basic terminology. 
A {\em linear code} of {\em length} $m$ and {\em dimension} $k$
over a finite field $\F_q$ is a $k$-dimensional subspace $C$ of 
the $m$-dimensional vector space $\F_q^m$; the elements of $C$ are 
called {\em codewords}. Such a code $C$ can be represented by 
a $k\times m$ {\em generator matrix} $G\in \F_q^{k\times m}$ of rank $k$, 
with the interpretation that any linear combination $y=xG$ with $x\in \F_q^k$ 
is a codeword of $C$. The \emph{support} of a codeword 
$y=(y_1,y_2,\ldots,y_m)\in C$ is the set 
$S(y)=\{i\in\{1,2,\ldots,m\}:y_i\neq 0\}$ of nonzero coordinates. 
For a nonempty set $Y\subseteq C$ of codewords, the \emph{combined} 
support is defined by $S(Y)=\cup_{y\in y} S(y)$. The combined support 
is {\em full} if $S(Y)=\{1,2,\ldots,m\}$.

Our two lower bounds use the following famous connection between the 
Tutte polynomial and code words of full combined support due to 
Crapo and Rota~\cite{CrapoR1970}:

\begin{Thm}[The Critical Theorem; Crapo and Rota~\cite{CrapoR1970}]
\label{thm:critical}
Let $d$ be a positive integer and let $C\subseteq\F_q^m$ be a linear code with
a generator matrix $G$. Then, the number of $d$-tuples of codewords in $C^d$
with full combined support is $(-1)^{\rho(G)}T_G\left(1-q^d,0\right)$.
\end{Thm}

Consider a linear code $C\subseteq\F_q^{m}$ with generator matrix $G$. 
Theorem~\ref{thm:critical} with $d=1$ implies that the number of codewords 
of $C$ with full support can be obtained as the evaluation of the Tutte
polynomial $T_G$ at a single point. Our proof of Theorem~\ref{thm:main-lb} 
will crucially rely on this connection. In essence, the property of the 
codeword $y=Gx$ having full support corresponds to $x$ being a solution of 
a system of linear homogeneous {\em inequations} 
$\alpha_1x_1+\alpha_2x_2+\ldots+\alpha_kx_k\neq 0$
over $\F_q$, one inequation for each column of~$G$. Geometrically, each such 
inequation can be viewed as a constraint that forces $x$ to lie not on 
a particular hyperplane through the origin, and a system of such constraints 
forces $x$ to lie properly inside a chamber of an arrangement of hyperplanes 
through the origin. 
The crux of our proof of Theorem~\ref{thm:main-lb} is to show via a sequence 
of lemmas that the task of computing the total volume of these hyperplane 
chambers is hard under \#ETH, even in the case when every hyperplane is 
defined by a vector with at most two nonzero entries; 
cf.~Lemma~\ref{lem:hies2-hard}.

Theorem~\ref{thm:main-lb} leaves open the complexity analysis for any fixed 
finite field. To establish hardness under \#ETH for any fixed finite field and
prove Theorem~\ref{thm:main-lb-fixed}, we will invoke 
Theorem~\ref{thm:critical} for larger values of the parameter~$d$ to access 
the codewords of full support in an extension code. In more precise terms, 
let $C\subseteq\F_q^m$ be a {\em base} code with generator matrix 
$G\in\F_q^{k\times m}$. For a positive integer $d$, we obtain the 
\emph{extension code} $\bar C\subseteq\F_{q^d}^m$ of the base code $C$ 
by embedding $G$ elementwise into $\F_{q^d}$ to obtain the generator matrix 
$\bar G\in\F_{q^d}^{k\times m}$ of $\bar C$. Theorem~\ref{thm:critical} 
applied to the base code $C$ with this $d$ implies that the number of 
codewords of the extension code $\bar C$ with full support can be obtained 
as the evaluation of the Tutte polynomial~$T_G$ of the base code at a single 
point. This is because for every $d$-tuple  
$(y^{(1)},y^{(2)},\ldots,y^{(d)})$ of codewords in $C^d$ with full combined 
support and $x^{(i)}G=y^{(i)}$ for $i=1,2,\ldots,d$, we can build a unique 
$\bar x=(\bar x_1,\ldots,\bar x_k)\in \F_{q^d}^k$ so that $\bar x\bar G$
is a codeword of $\bar C$ with full support. Indeed, $\F_{q^d}$ can be 
represented as the polynomial quotient ring $\F_{q}[w]/\bra I(w)\ket$
in the indeterminate $w$, where $I(w)\in\F_q[w]$ is an irreducible polynomial 
of degree $d$ over $\F_q$, and we can build the scalars 
$\bar x_j\in\F_{q^d}$ in this representation as 
$\bar x_j=\sum_{i=0}^{d-1} x_j^{(i)}w^i$ for $j=1,2,\ldots,d$. 
This representation also shows that the reverse transform is possible: 
from every codeword of full support in $\bar C$, we can construct 
a unique $d$-tuple of codewords in $C^d$ with full combined support. 
Hence, their cardinalities are the same. We state this well-known connection 
as a lemma.

\begin{Lem}[Counting codewords of full support in extension code]
\label{lem:counting-extension-full-support}
There is a one-to-one correspondence between codewords of full support
in $\bar C\subseteq\F_{q^d}^m$ and $d$-tuples of codewords from
$C\subseteq\F_{q}^m$ having full combined support.
\end{Lem}

Thus, we can rely on a Tutte
polynomial of the generator matrix of the {\em base} code $C$ to access 
the count of full-support codewords for the extension code $\bar C$. 
In particular, the base
code can be over any fixed finite field, including the binary field, which 
enables establishing hardness under \#ETH for any fixed finite field. 
The crux of our proof of 
Theorem~\ref{thm:main-lb-fixed} is to establish hardness under \#ETH for 
systems of linear homogeneous {\em sum}-inequations 
$\alpha_1x_1+\alpha_2x_2+\ldots+\alpha_kx_k\neq 0$ with 
$\alpha_i\in\{-1,0,1\}$ for all $i=1,2,\ldots,k$, even in the case when 
$\alpha_i\neq 0$ for at most three $i$. In particular, sum-inequations 
are representable over any fixed finite field, which enables our hardness 
reductions under \#ETH as a sequence of lemmas culminating in 
Theorem~\ref{thm:main-lb-fixed}.

Let us conclude this section with a brief discussion of related work and
techniques. First, our combinatorial techniques on instances of constraint
satisfaction problems are influenced by earlier hardness results,
such as the seminal work of Traxler~\cite{Traxler2008}. Similarly, 
the work of Kowalik and Socala~\cite{KowalikS2014} demonstrates how to 
bridge between combinatorial and sparse algebraic constraints in the form
of generalized list colorings. Earlier work on $n^{o(n)}$-form tight lower
bounds under ETH includes e.g.~the work of 
Cygan~{\em et al.}~\cite{CyganFGKMPS2017} on graph embedding problems. A more recent example
is the work of Fomin, Lokshtanov, Mihajlin, Saurabh, and Zehavi on the Hadwiger 
number and related contraction problems~\cite{FominLMSZ2020}.

Finally, our present focus is on tuples of codewords of full support 
in a linear code via Theorem~\ref{thm:critical}; dually, words of least
positive support size determine the minimum distance of the code, a quantity
which is also known to be hard to compute; cf.~Vardy~\cite{Vardy1997}.

\subsection{Organization}
The rest of this paper is organized as follows. Section~\ref{sect:lb}
proves our main lower-bound theorems, Theorem~\ref{thm:main-lb} and
Theorem~\ref{thm:main-lb-fixed}. Sections~\ref{sect:ub} and \ref{sect:ub-wt2} 
present our upper-bound algorithm designs for Theorems~\ref{thm:main-ub}
and \ref{thm:main-ub-wt2}, respectively.

\section{Lower bounds}
\label{sect:lb}

This section proves our two main lower-bound theorems,
Theorem~\ref{thm:main-lb} and Theorem~\ref{thm:main-lb-fixed}.
We start with preliminaries on constraint satisfaction problems, the
counting exponential time hypothesis and sparsification, and then proceed
to develop the technical preliminaries and tools needed to transform 
combinatorial CSP instances into appropriately restricted algebraic versions
that can then be accessed in a coding-theoretic context.

\subsection{Constraint satisfaction problems}

For nonnegative integers $d$, $a$, $v$, and $m$,
a {\em constraint satisfaction problem instance} $\varphi$ with parameters
$(d,a,v,m)$---or briefly, a $(d,a,v,m)$-{\em CSP instance}---consists
of $v$ {\em variables} $x_1,x_2,\ldots,x_v$ and
$m$ {\em constraints} $C_1,C_2,\ldots,C_m$ such that
\begin{enumerate}
\item
associated with each variable $x_i$, there is an at-most-$d$-element set $D_i$,
the {\em domain} of $x_i$; and
\item
associated with each constraint $C_j$, there is an $a$-tuple
$S_j=(x_{j_1},x_{j_2},\ldots,x_{j_a})$ of distinct variables as well as
a set $P_j\subseteq D_{j_1}\times D_{j_2}\times\cdots\times D_{j_a}$
of {\em permitted combinations of values} for the variables.
\end{enumerate}
We say that the parameter $d$ is the {\em domain size} of the variables
and the parameter $a$ is the {\em arity} of the constraints.
We may omit the parameters $v$ and $m$ and simply refer to a $(d,a)$-CSP
instance if this is convenient.

We say that a $(d,a,v,m)$-CSP instance $\varphi$ is {\em satisfiable}
if there exists a {\em satisfying assignment}
$w\in D_1\times D_2\times\cdots\times D_v$
such that for every $j=1,2,\ldots,m$ it holds that $w$ assigns a
permitted combination of values to the constraint $C_j$---that is---we
have $(w_{j_1},w_{j_2},\ldots,w_{j_a})\in P_j$; otherwise, we say
that $\varphi$ is {\em unsatisfiable}. Let us write 
$\SAT(\varphi)\subseteq D_1\times D_2\times \cdots \times D_v$ for the
set of all satisfying assignments of $\varphi$.

Let us write $(d,a,v,m)$-CSP for the task of deciding whether a given
$(d,a,v,m)$-CSP instance is satisfiable. Similarly, let us write 
$\#(d,a,v,m)$-CSP for the task of counting the number of satisfying
assignments to a given $(d,a,v,m)$-CSP instance.

A constraint where all but one combination of values to the variables
is permitted is called a {\em clause}. Instances consisting of
clauses over variables with a binary domain are said to be in
{\em conjunctive normal form} (CNF). We refer to instances in CNF with
arity $k$ as $k$-CNF, where the parameter $k$ is the {\em length} of
the clauses.

\subsection{The counting exponential-time hypothesis and sparsification}

\label{sect:num-eth}

No efficient algorithm is known for solving constraint satisfaction problems 
in the general case. As such, we will establish our present hardness results 
under the following hypothesis of
Dell~{\em et al.}~\cite{DellHMTW2014}, which relaxes the Exponential
Time Hypothesis of Impagliazzo and Paturi~\cite{ImpagliazzoP1999}.

\begin{Hyp}[Counting exponential time hypothesis (\#ETH); Dell~{\em et al.}~\cite{DellHMTW2014}]
\label{hyp:num-eth}
There exists a constant $c>0$ such that there is no deterministic 
algorithm that solves a given $n$-variable instance of $\#3$-CNF 
in time $\exp(cn)$.
\end{Hyp}

We will also need a counting-variant of the Sparsification Lemma 
of Impagliazzo, Paturi, and Zane~\cite{ImpagliazzoPZ2001} due to 
Dell~{\em et al.}~\cite{DellHMTW2014} 
(see also Flum and Grohe~\cite{FlumG2006}).

\begin{Lem}[Counting sparsification; Dell~{\em et al.}~\cite{DellHMTW2014}]
\label{lem:counting-sparsification}
For $k\geq 2$, there exists a computable function $\sigma:\Nat^2\rightarrow\Nat$
and a deterministic algorithm that, for $p\in\Nat$ and an $n$-variable 
$\#k$-CNF instance $\varphi$ given as input, in time $O\bigl(t\cdot \poly n\bigr)$ 
computes $\#k$-CNF instances $\varphi_1,\varphi_2,\ldots,\varphi_t$, each
over the same variables and variable domains as $\varphi$, 
such that 
\begin{enumerate}
\item
$t\leq 2^{n/p}$;
\item
$\SAT(\varphi)=\cup_{i=1}^{t}\SAT(\varphi_i)$ where the union consists of disjoint sets; and
\item
each variable occurs in at most $\sigma(k,p)$ clauses of $\varphi_i$. 
\end{enumerate}
\end{Lem}
  
\subsection{Hardness of bipartite CSPs}

It will be convenient to base our main hardness reductions on CSPs whose
constraints have the topology of a bipartite graph. Towards this end, 
this section presents variants of well-known 
(e.g.~cf.~Traxler~\cite{Traxler2008}) hardness reductions that have been 
modified to establish hardness in the bipartite case. 

In more precise terms, let us study a CSP instance with arity $a=2$. 
It is immediate that we can view the constraints of such an instance as 
the edges of a (directed) graph whose vertices correspond to the variables
of the instance. We say that such a CSP instance is {\em bipartite} if 
this graph is bipartite. 

\begin{Lem}[Hardness of bipartite \#CSP under \#ETH]
\label{lem:eth-hard-bip-csp}
Assuming $\#ETH$,
there is a constant $b>0$ such that there is no deterministic algorithm 
that solves a given bipartite $\#(8,2,v,O(v))$-CSP instance in time $\exp(bv)$.
\end{Lem}
\begin{Proof}
Let $c$ be the constant in Hypothesis~\ref{hyp:num-eth} and
let $\varphi$ be a $n$-variable instance of $\#3$-CNF.
Select a positive integer $p$ so that $p>2/c$. 
Run the sparsification algorithm in Lemma~\ref{lem:counting-sparsification}
on $\varphi$ to obtain in time $O(2^{cn/2}\poly n)$ the
\#$3$-CNF instances $\varphi_1,\varphi_2,\ldots,\varphi_t$ 
with $t\leq 2^{cn/2}$. 

Let us transform $\varphi_i$ into a bipartite $\#(2^3,2)$-CSP instance
$\varphi_i'$ with $|\SAT(\varphi_i')|=|\SAT(\varphi_i)|$. 
Without loss of generality we may assume that every variable occurs in 
at least one clause.
Let us assume that $\varphi_i$ consists of $m$ clauses $C_1,C_2,\ldots,C_m$ 
over $n$ variables $x_1,x_2,\ldots,x_n$ with domains $D_1,D_2,\ldots,D_n$,
respectively.
By Lemma~\ref{lem:counting-sparsification}, we have $m\leq\sigma(3,p)n=O(n)$.
Let us write $(x_{j_1},x_{j_2},x_{j_3})$ for the support of $C_j$
and $P_j\subseteq D_{j_1}\times D_{j_2}\times D_{j_3}$ for the permitted
values of $C_j$.

The construction of $\varphi_i'$ is as follows.
For each clause $C_j$ with $j=1,2,\ldots,m$, introduce a variable $C_j'$ 
with domain $D_{j_1}\times D_{j_2}\times D_{j_3}$ into $\varphi_i'$. 
For each variable $x_j$ with $j=1,2,\ldots,n$, introduce
a variable $x_j'$ with domain $D_j$ into $\varphi_i'$.
For each clause $C_j$ with $j=1,2,\ldots,m$ and each $\ell=1,2,3$, 
introduce a constraint with support $(x_{j_\ell}',C_j')$ and 
permitted combinations 
$P'_{j,\ell}=\{(w,(w_1,w_2,w_3))\in D_{j_\ell}\times P_j:w=w_\ell\}
\subseteq D_{j_\ell}\times (D_{j_1}\times D_{j_2}\times D_{j_3})$
into $\varphi_i'$. In total $\varphi_i'$ thus has 
$v\leq(\sigma(3,p)+1)n$ variables and $3m\leq 3\sigma(3,p)n=O(v)$ constraints. 
It is also immediate that $\varphi_i'$ has domain size $2^3$, arity $2$, and 
bipartite structure as a graph. Furthermore, since every variable of 
$\varphi_i$ occurs in at least one clause, it is immediate that there is
a one-to-one correspondence between $\SAT(\varphi_i)$ and $\SAT(\varphi_i')$.
The transformation from $\varphi_i$ to $\varphi_i'$ is clearly computable
in time $\poly n$. 

To reach a contradiction, suppose now that there is a deterministic algorithm 
that solves a given bipartite $\#(2^3,2,v,O(v))$-CSP instance in time 
$\exp(bv)$ for a constant $b>0$ with $b<c/(2(\sigma(3,p)+1))$. Then, we could 
use this algorithm to solve each of the $t\leq 2^{cn/2}$ instances $\varphi_i'$ 
for $i=1,2,\ldots,t$ in total time $\exp(c'n)$ for a constant $c'<c$. 
But since $|\SAT(\varphi_i')|=|\SAT(\varphi_i)|$, this means that we
could solve each of the instances $\varphi_i$, and thus the \#3-CNF instance 
$\varphi$ by Lemma~\ref{lem:counting-sparsification}, in similar total time, 
which contradicts Hypothesis~\ref{hyp:num-eth}.
\end{Proof}

The next lemma contains a well-known tradeoff (e.g.~\cite{Curticapean2018}) 
that amplifies the lower bound on the running time by enlarging the domains 
of the variables. 

\begin{Lem}[Hardness amplification by variable aggregation under \#ETH]
\label{lem:quasipoly-hard}
Assuming $\#$ETH,
there is no deterministic algorithm that solves a given 
bipartite $\#(\lfloor\sqrt{n}\rfloor,2,n,O(n\polylog n))$-CSP 
instance in time $n^{o(n)}$.
\end{Lem}
\begin{Proof}
We establish hardness via Lemma~\ref{lem:eth-hard-bip-csp}.
Let $\varphi$ be a bipartite $\#(8,2,v,O(v))$-CSP instance.
Without loss of generality---by padding with extra variables constrained
to unique values---we may assume that (i) the variables of $\varphi$
are $x_1,x_2,\ldots,x_v,y_1,y_2,\ldots,y_v$, (ii) every constraint
of $\varphi$ has support of the form $(x_i,y_j)$ for some $i,j=1,2,\ldots,v$,
and (iii) $v\geq 2$. Let $\epsilon>0$ be a constant whose value is fixed 
later and let $g=\lceil \epsilon\log v\rceil$. Group the variables 
$x_1,x_2,\ldots,x_v$ into pairwise disjoint sets 
$X_1,X_2,\ldots,X_{\lceil v/g\rceil}$ of at most $g$ variables each. 
Similarly, group the variables $y_1,y_2,\ldots,y_v$
into pairwise disjoint sets $Y_1,Y_2,\ldots,Y_{\lceil v/g\rceil}$ of at most
$g$ variables each. 

Let us construct from $\varphi$ a bipartite $\#$CSP instance $\varphi'$ with 
$|\SAT(\varphi)|=|\SAT(\varphi')|$ as follows. The variables of $\varphi'$ 
are $X_1,X_2,\ldots,X_{\lceil v/g\rceil}$ 
and $Y_1,Y_2,\ldots,Y_{\lceil v/g\rceil}$ so that the domain of each variable
is the Cartesian product of the domains of the underlying variables 
of $\varphi$. The constraints of $\varphi'$ are obtained by extension 
of the constraints of $\varphi$ as follows. For each constraint with
support $(x_i,y_j)$ in $\varphi$, let $i'$ and $j'$ be the unique indices
with $x_i\in X_{i'}$ and $y_j\in Y_{j'}$, and introduce a constraint
with support $(X_{i'},Y_{j'})$ into $\varphi'$; set the permitted values of 
this constraint so that they force a permitted value to the variables $x_i$ 
and $y_j$ as part of the variables $X_{i'}$ and $Y_{j'}$ but otherwise
do not constrain the values of $X_{i'}$ and $Y_{j'}$. This completes
the construction of $\varphi'$. It is immediate that $\varphi'$ is bipartite
and that $|\SAT(\varphi)|=|\SAT(\varphi')|$ holds. Furthermore, $\varphi'$ 
has $n=2\lceil v/\lceil\epsilon\log v\rceil\rceil$ variables, each with 
domain size at most $8^{\lceil\epsilon\log v\rceil}$, and $O(v)$ constraints;
that is, $O(n\polylog n)$ constraints. 
Choosing $\epsilon=1/7$, we have $8^{\lceil\epsilon\log v\rceil}\leq\sqrt{n}$ 
for all large enough $n$. The transformation from $\varphi_i$ to 
$\varphi_i'$ is clearly computable in time $\poly v$. 

To reach a contradiction, suppose now that there is a deterministic algorithm 
that solves a given bipartite 
$\#(\lfloor\sqrt{n}\rfloor,2,n,O(n\polylog n))$-CSP 
instance in time $n^{o(n)}=\exp(o(n\log n))$. Then, we could 
use this algorithm to solve $\varphi'$, and hence $\varphi$ by 
$|\SAT(\varphi')|=|\SAT(\varphi)|$, in time $\exp(o(v))$,
which contradicts Lemma~\ref{lem:eth-hard-bip-csp}.
\end{Proof}

\subsection{Linear inequation systems and chambers of hyperplane arrangements}

We are now ready to introduce our main technical tool, namely CSPs over 
finite fields whose constraints are of a special geometric form. 
(For preliminaries on finite fields, 
 cf.~e.g.~Lidl and Niederreiter~\cite{LidlN1983}.)
More precisely, let us write $\F_q$ for the finite field with $q$ elements,
$q$ a prime power, and let $x_1,x_2,\ldots,x_n$ be variables taking values
in $\F_q$. For $\alpha_1,\alpha_2,\ldots,\alpha_n\in\F_q$,
$\beta\in\F_q$, and $S=\{j\in\{1,2,\ldots,n\}:\alpha_j\neq 0\}$, 
we say that the constraint
\begin{equation}
\label{eq:off-hyperplane}
\alpha_1x_1+\alpha_2x_2+\ldots+\alpha_n x_{n}\neq \beta
\end{equation}
is a ({\em linear}) {\em inequation} of arity (or {\em weight}) $|S|$. 
We say that the inequation is {\em homogeneous} if $\beta=0$ and 
{\em inhomogeneous} otherwise. We say that the inequation is 
a {\em sum-inequation} if for all $j\in S$ we have $\alpha_j\in\{1,-1\}$.

Previously, the complexity of inequations of low arity has been 
studied for example by Kowalik and Socala~\cite{KowalikS2014} under 
the terminology of generalized list colorings of graphs. We also remark
that for $|S|\geq 1$ one can view \eqref{eq:off-hyperplane} geometrically 
as the constraint that a point $x\in\F_q^n$ does not lie in the hyperplane 
defined by the coefficients $\alpha_1,\alpha_2,\ldots,\alpha_n$ and $\beta$; 
accordingly, a system of constraints of this form is satisfied by a point $x$ 
if and only if $x$ lies properly 
inside a chamber of the corresponding hyperplane arrangement, and the
task of counting the number of such points corresponds to determining
the total volume of the chambers in $\F_q^n$.
(Cf.~Orlik and Terao~\cite{OrlikT1992}, Dimca~\cite{Dimca2017}, and
Aguiar and Mahajan~\cite{AguiarM2017} for hyperplane arrangements.)

Here our objective is to establish that systems of inequations are hard 
to solve under \#ETH already in the homogeneous case and for essentially 
the smallest nontrivial arity, using our preliminaries on bipartite CSPs 
to enable the hardness reductions. We start with inequations of arity two 
in the following section, and proceed to sum-inequations of arity three 
in the next section.

\subsection{Homogeneous inequation systems of arity two}

Our first goal is to show that counting the number of solutions to a 
homogeneous inequation system of arity two over a large-enough 
field is hard under \#ETH. 

It will be convenient to start by establishing
hardness of {\em modular} constraints of arity two, and then proceed to
the homogeneous case over $\F_q$ by relying on the cyclic structure
of the multiplicative group of $\F_q$. The modular
setting will also reveal the serendipity of our work with bipartite CSPs. 
Towards this end, let $x_1,x_2,\ldots,x_n$, $y_1,y_2,\ldots,y_n$, and $z$ 
be $2n+1$ variables taking values in $\Z_M$, the integers modulo $M$. 
We say that an inequation of arity two over $\Z_M$ 
is {\em special modular} if it is one of the following forms:
(i) $x_i-y_j\neq c$, (ii) $x_i-z\neq c$, or (iii) $y_j-z\neq c$ for
$i,j=1,2,\ldots,n$ and $c\in\Z_M$. A CSP instance over $\Z_M$ 
is {\em special modular} if all of its constraints are special modular.

\begin{Lem}[Hardness of special modular systems under \#ETH]
\label{lem:special-hard}
Assuming $\#$ETH,
there is no deterministic algorithm that in time $n^{o(n)}\poly M$ 
solves a given special modular \#$(M,2,2n+1,O(Mn\polylog n))$-CSP instance 
over $\Z_M$ with $M\geq 3n$.
\end{Lem}
\begin{Proof}
We establish hardness via Lemma~\ref{lem:quasipoly-hard}.
Let $\varphi$ be a bipartite 
$\#(\lfloor\sqrt{n}\rfloor,2,n,O(n\polylog n))$-CSP instance.
Without loss of generality---by padding with extra variables constrained
to unique values---we may assume that the variables of $\varphi$
are $x_1,x_2,\ldots,x_n,y_1,y_2,\ldots,y_n$ and every constraint
of $\varphi$ has support of the form $(x_i,y_j)$ for some $i,j=1,2,\ldots,n$.
Furthermore, by relabeling of the domains as necessary, we can assume that
all variables $x_i$ have domain $\{d,2d,\ldots,d^2\}$ and all variables
$y_j$ have domain $\{0,1,\ldots,d-1\}$ with $d=\lfloor\sqrt{n}\rfloor$.
Let us now construct a special modular CSP instance $\varphi'$ as follows. 
Let $M\geq 3n\geq 3d^2$. Introduce the $2n+1$ 
variables $x_1,x_2,\ldots,x_n$, $y_1,y_2,\ldots,y_n$, and $z$ into 
$\varphi'$ so that each variable has domain $\Z_M=\{0,1,\ldots,M-1\}$.
For each $i=1,2,\ldots,n$, 
force $x_i\in\{z+d,z+2d,\ldots,z+d^2\}$ modulo $M$
by introducing $M-d$ special modular constraints of type (ii) into $\varphi'$.
For each $j=1,2,\ldots,n$, 
force $y_j\in\{z,z+1,\ldots,z+d-1\}$ modulo $M$
by introducing $M-d$ special modular constraints of type (iii) into $\varphi'$.
We observe that the introduction of these constraints into $\varphi'$ 
forces that for all $i,j=1,2,\ldots,n$ we have 
$x_i-y_j\in\{1,2,\ldots,d^2\}$ modulo $M$, and the values
of $x_i$ and $y_j$ modulo $M$ are uniquely determined by the 
difference $x_i-y_j$ modulo $M$. 
Finally, for each constraint of $\varphi$ with support of the form 
$(x_i,y_j)$ for some $i,j=1,2,\ldots,n$, use at most $M$ 
special modular constraints of type (i) to force the values of $x_i$ and $y_j$ 
to the permitted pairs of values.
It is immediate that $|\SAT(\varphi')|=M|\SAT(\varphi)|$; indeed, each
satisfying assignment to $\varphi$ corresponds to exactly $M$ satisfying
assignments to $\varphi'$, one for each possible choice of value to $z$.
Furthermore, $\varphi'$ is computable from $\varphi$ in time 
$\textrm{poly}(M,n)$.
We also observe that $\varphi'$ has $2n+1$ variables, $O(Mn\polylog n)$
constraints, domain size $3n$, and arity $2$.

To reach a contradiction, suppose now that there is a deterministic algorithm 
that in time $n^{o(n)}\poly M$ solves a given  special modular 
\#$(M,2,2n+1,O(Mn\polylog n))$-CSP instance over $\Z_M$ with $M\geq 3n$.
Then, we could use this algorithm to solve $\varphi'$, and hence $\varphi$ by 
$|\SAT(\varphi')|=M|\SAT(\varphi)|$, in time $n^{o(n)}$,
which contradicts Lemma~\ref{lem:quasipoly-hard}.
\end{Proof}

We are now ready to establish hardness of homogeneous inequation
systems of arity two over $\F_q$ for large-enough $q$. For arithmetic 
in $\F_q$, we tacitly assume an appropriate irreducible polynomial and 
a generator $\gamma$ for the multiplicative group of $\F_q$ are supplied 
as part of the input. (For algorithmics for finite fields, 
cf.~e.g.~von zur Gathen and Gerhard~\cite{vonzurGathenG2013}.)

\begin{Lem}[Hardness of homogeneous inequation systems of arity two under \#ETH]
\label{lem:hies2-hard}
Assuming $\#$ETH,
there is no deterministic algorithm that in time $n^{o(n)}\poly q$ solves 
a given \#$(q,2,2n+1,O(qn\polylog n))$-CSP instance with the structure of 
a homogeneous inequation system over $\F_q$ with $q\geq 3n+1$.
\end{Lem}
\begin{Proof}
We proceed via Lemma~\ref{lem:special-hard}.
Let $\varphi$ be a special modular \#$(M,2,2n+1,O(Mn\polylog n))$-CSP instance
with variables $x_1,x_2,\ldots,x_n,y_1,y_2,\ldots,y_n,z$ taking values
in $\Z_M$ for $M\geq 3n$. By choosing a large enough $M$ in 
Lemma~\ref{lem:special-hard}, we may assume that $M+1$ is a prime power.
Let us construct a homogeneous inequation system $\varphi'$ 
over $\F_q$ with $q=M+1$ as follows. Let 
$\gamma$ be a generator for the multiplicative group of $\F_q$. 
Introduce into $\varphi'$ the variables 
$x_1',x_2',\ldots,x_n'$, $y_1',y_2',\ldots,y_n'$, and $z'$, each 
taking values in $\F_q$. Introduce the homogeneous inequations
$x_i'\neq 0$, $y_j'\neq 0$, and $z'\neq 0$ for all $i,j=1,2,\ldots,n$
into $\varphi'$. By the cyclic structure of the multiplicative group
of $\F_q$, we have that to arbitrary {\em nonzero} values of the variables
$x_i'$, $y_j'$, $z'$ in $\F_q$, there correspond unique 
integers $x_i$, $y_j$, $z$ modulo $q-1$ such that 
$x_i'=\gamma^{x_i}$, $y_j'=\gamma^{y_j}$, 
and $z'=\gamma^{z}$ for all $i,j=1,2,\ldots,n$. 
Furthermore, under this correspondence, each special modular constraint
$x_i-y_j\neq c$ over $\Z_{M}=\Z_{q-1}$ corresponds to the 
homogeneous inequation $x_i'-\gamma^{c}y_j'\neq 0$
of arity $2$ over~$\F_q$; indeed, we have 
$x_i'-\gamma^cy_j'\neq 0$ iff
$x_i'\neq\gamma^cy_j'$ iff
$\gamma^{x_i}\neq\gamma^{c+y_j}$ iff
$x_i\neq c+y_j$ modulo $q-1$ iff 
$x_i-y_j\neq c$ modulo $q-1$.
The special modular constraints $x_i-z\neq c$ and $y_j-z\neq c$ have
similar correspondence with homogeneous inequations 
$x_i'-\gamma^{c}z'\neq 0$ and $y_j'-\gamma^{c}z'\neq 0$, respectively.
We can thus complete the construction of
$\varphi'$ by inserting the constraints corresponding to the 
constraints of $\varphi$ into $\varphi'$; in particular, we have
$|\SAT(\varphi)|=|\SAT(\varphi')|$. The transformation from $\varphi$
to $\varphi'$ is clearly computable in time $\textrm{poly}(n,q)$. 
It thus follows from 
Lemma~\ref{lem:special-hard} that, assuming \#ETH,
there is no deterministic algorithm that in time $n^{o(n)}\poly q$
solves a given \#$(q,2,2n+1,O(qn\polylog n))$-CSP instance with the 
structure of a homogeneous inequation system over $\F_q$ with $q\geq 3n+1$.
\end{Proof}

\subsection{Homogeneous sum-inequation systems of arity three}

We now proceed to look at homogeneous inequation systems 
with $\{-1,0,1\}$-coefficients on the variables; that is, we establish 
under \#ETH the hardness of counting the number of solutions to 
a homogeneous {\em sum}-inequation system of low arity. 
Bipartiteness in the input of the reduction 
will again be serendipitous in achieving low arity. 
In particular, bipartiteness will enable us to reduce to a system of 
homogeneous sum-inequations of arity three whose solvability in relation to
the original system can be established via the existence of Sidon sets.

For an Abelian group~$A$, we say that a subset $S\subseteq A$ is 
a {\em Sidon set} if for any $x,y,z,w\in S$ of which at least three are 
different, it holds that $x+y\neq z+w$. An Abelian group is {\em elementary}
Abelian if all of its nontrivial elements have order $p$ for a prime $p$.
The additive group of a finite field $\F_q$ is elementary Abelian.

\begin{Lem}[Existence of Sidon sets; 
{Babai and S\'os~\cite[Corollary~5.8]{BabaiS1985}}]
\label{lem:sidon}
Elementary Abelian groups of order $q$ have Sidon sets of size $q^{1/2+o(1)}$.
\end{Lem}

We are now ready for the main result of this section. 

\begin{Lem}[Hardness of homogeneous sum-inequation systems of arity three under \#ETH]
\label{lem:hsies2-hard}
Assuming $\#$ETH,
there is no deterministic algorithm that in time $n^{o(n)}\poly q$
solves a given \#$(q,3,2(n+q),O(q^2\polylog q))$-CSP instance with 
the structure of a homogeneous sum-inequation system over $\F_q$ 
with $q\geq n^{1+o(1)}$.
\end{Lem}
\begin{Proof}
We proceed via Lemma~\ref{lem:quasipoly-hard}.
Let $\varphi$ be a bipartite 
$\#(\lfloor\sqrt{n}\rfloor,2,n,O(n\polylog n))$-CSP instance.
Without loss of generality---by padding with extra variables constrained
to unique values---we may assume that the variables of $\varphi$
are $x_1,x_2,\ldots,x_n,y_1,y_2,\ldots,y_n$ and every constraint
of $\varphi$ has support of the form $(x_i,y_j)$ for some $i,j=1,2,\ldots,n$.
Furthermore, by relabeling of the domains as necessary, we can assume that
all variables $x_i$ and $y_j$ have domain $\{1,2,\ldots,d\}$ 
with $d=\lfloor\sqrt{n}\rfloor$.

Let us construct a homogeneous sum-inequation system $\varphi'$ 
over $\F_q$ with $q\geq n$ as follows. 
Introduce the variables 
$x_1',x_2',\ldots,x_n'$,
$y_1',y_2',\ldots,y_n'$, 
$s_1',s_2',\ldots,s_d'$, 
$t_1',t_2',\ldots,t_d'$, 
$r_1',r_2',\ldots,r_{q-2d}'$, 
and 
$v_1',v_2',\ldots,v_q'$,
each taking values over $\F_q$, into $\varphi'$.
In total there are thus $2(n+q)$ variables.

We introduce six different types of homogeneous sum-inequations 
into $\varphi'$. Let $g:\{1,2,\ldots,d\}^2\rightarrow\{1,2,\ldots,q\}$ be 
an arbitrary but fixed injective map.

First, inequations of type (i) force the $q$ variables
$s_1',s_2',\ldots,s_d'$, $t_1',t_2',\ldots,t_d'$, $r_1',r_2',\ldots,r_{q-2d}'$
to take pairwise distinct values; this can be forced with 
$q(q-1)/2$ homogeneous sum-inequations of arity $2$.

Second, 
inequations of type (ii) force the $q$ variables
$v_1',v_2',\ldots,v_q'$ to take pairwise distinct values; this can 
be forced with $q(q-1)/2$ homogeneous sum-inequations of arity $2$.

Third, for each $a,b\in\{1,2,\ldots,d\}$, we force the equality
$s_a'+t_b'=v_{g(a,b)}'$ by introducing
$q-1$ homogeneous sum-inequations $s_a'+t_b'-v_k'\neq 0$---let us call these
inequations of type (iii)---one inequation for each 
$k\in\{1,2,\ldots,q\}\setminus\{g(a,b)\}$.

Fourth, inequations of type (iv) force the $n$ variables 
$x_i'$ to take values in the set of values of the 
variables $s_1',s_2',\ldots,s_d'$; together with (i), 
this can be forced with homogeneous sum-inequations 
$x_i'-t_b'\neq 0$ and $x_i'-r_\ell'\neq 0$ for all $i=1,2,\ldots,n$, 
$b=1,2,\ldots,d$, and $\ell=1,2,\ldots,q$.

Fifth, inequations of type (v) force the $n$ variables 
$y_j'$ to take values in the set of values of the 
variables $t_1',t_2',\ldots,t_d'$; together with (i), 
this can be forced with homogeneous sum-inequations 
$y_j'-s_a'\neq 0$ and $y_j'-r_\ell'\neq 0$ for all $j=1,2,\ldots,n$, 
$b=1,2,\ldots,d$, and $\ell=1,2,\ldots,q$.

Sixth, for each constraint with support $(x_i,y_j)$ in $\varphi$
for some $i,j=1,2,\ldots,n$, and letting $P\subseteq\{1,2,\ldots,d\}^2$ be the
set of permitted values for the constraint, introduce 
the homogeneous sum-inequations
$x_i'+y_j'-v_k'\neq 0$ for each $k\in\{1,2,\ldots,q\}\setminus g(P)$;
let us call these inequations of type (vi).

This completes the transformation from $\varphi$ to $\varphi'$, which is 
clearly computable in time $\poly(n,q)$. We observe that $\varphi'$ has
domain size $q$, arity $3$, $2(n+q)$ variables, and $O(q^2\polylog q)$ 
constraints. 

Next we claim that for all large enough $q$ we have 
$|\SAT(\varphi')|=f(q,d)\cdot|\SAT(\varphi)|$
for a positive-integer-valued function $f(q,d)$ of the parameters $q,d$.
Indeed, let $f(q,d)$ be the total number of solutions to 
the system of inequations consisting of the variables 
$s_1',s_2',\ldots,s_d'$, $t_1',t_2',\ldots,t_d'$, $r_1',r_2',\ldots,r_{q-2d}'$,
$v_1',v_2',\ldots,v_q'$ and all the inequations of types (i), (ii), and (iii).
Recalling that $q\geq n^{1+o(1)}\geq d^{2+o(1)}$, 
from Lemma~\ref{lem:sidon} we have that for all large enough $q$ the 
additive group of $\F_q$ contains a Sidon set of size $2d$. Assign each 
element of this Sidon set to exactly one of the variables 
$s_1',s_2',\ldots,s_d',t_1',t_2',\ldots,t_d'$ to 
conclude that the sums $s_a'+t_b'$ are distinct for all $a,b=1,2,\ldots,d$. 
Assign the remaining variables to distinct values in one of the 
$(q-2d)!(q-d^2)!$ possible ways to conclude that $f(q,d)\geq 1$. 
Fix one of the $f(q,d)$ solutions.
Inequations of type (iv) are by definition satisfied if and only
if for all $i=1,2,\ldots,n$ we have that 
$x_i'$ takes a value in the set of values for $s_1',s_2',\ldots,s_d'$.
Similarly, 
inequations of type (v) are by definition satisfied if and only
if for all $j=1,2,\ldots,n$ we have that 
$y_j'$ takes a value in the set of values for $t_1',t_2',\ldots,t_d'$.
Consider any such assignment to $x_i'$ and $y_j'$ for $i,j=1,2,\ldots,n$. 
Suppose that $x_i'=s_a'$ and $y_j'=t_b'$ for $a,b=1,2,\ldots,d$. Then,
$x_i'+y_j'=s_a'+t_b'=v_{g(a,b)}'$ since inequations of type (iii) are
satisfied. Suppose now $\varphi$ has a constraint with support $(x_i,y_j)$
and permitted values $P\subseteq\{1,2,\ldots,d\}^2$. By construction, 
we have that the inequations of type (vi) originating from this constraint 
are satisfied {\em if and only if} $(a,b)\in P$. Thus, 
we have $|\SAT(\varphi')|=f(q,d)\cdot|\SAT(\varphi)|$ as claimed.

To reach a contradiction, suppose that there is a deterministic algorithm 
that in time $n^{o(n)}\poly q$ solves a given 
\#$(q,3,2(n+q),O(q^2\polylog q))$-CSP instance with the structure of 
a homogeneous sum-inequation system over $\F_q$ with $q\geq n^{1+o(1)}$.
Let $\varphi$ be a bipartite 
$\#(\lfloor\sqrt{n}\rfloor,2,n,O(n\polylog n))$-CSP instance and take
$q=n^{1+o(1)}$.
First, use the assumed algorithm to the system of inequations consisting of 
the variables 
$s_1',s_2',\ldots,s_d'$, $t_1',t_2',\ldots,t_d'$, $r_1',r_2',\ldots,r_{q-2d}'$,
$v_1',v_2',\ldots,v_q'$ and all the inequations of types (i), (ii), and (iii).
The algorithm returns $f(q,d)$ as the solution.
Then, construct $\varphi'$ from $\varphi$ and use the algorithm on $\varphi'$ 
to get $|\SAT(\varphi')|$ as the solution. Divide by $f(q,d)$ to obtain
$|\SAT(\varphi)|$. Since the total running time
is $n^{o(n)}$, we obtain a contradiction to Lemma~\ref{lem:quasipoly-hard}.
\end{Proof}

We are now ready to complete our proofs of 
Theorem~\ref{thm:main-lb} and Theorem~\ref{thm:main-lb-fixed}.

\subsection{Proof of Theorem~\ref{thm:main-lb}}
We will rely on Lemma~\ref{lem:hies2-hard} and Theorem~\ref{thm:critical}. 
Let $\varphi$ be \#$(q,2,2n+1,O(qn\polylog n))$-CSP instance with the 
structure of a homogeneous inequation system over $\F_q$ with $q=3n+1$.
Take $k=2n+1$ and construct a $k\times m$ matrix $G\in\F_q^{k\times m}$ so
that each column of $G$ corresponds to a unique homogeneous
inequation of $\varphi$; in particular, every column of $G$
has at most two nonzero entries.
For all $x\in\F_q^k$ we have that $xG$ has full support 
if and only if $x\in\SAT(\varphi)$. Theorem~\ref{thm:critical} with $d=1$ 
thus implies that $(-1)^{\rho(G)}T_G(1-q,0)=|\SAT(\varphi)|$.
Since $m=O(qn\polylog n)$, we have $m=k^{O(1)}$. Furthermore, $q=k^{O(1)}$.
An algorithm that computes the Tutte polynomial $T_G$ in time $k^{o(k)}$
would thus enable us to compute $|\SAT(\varphi)|$ in time $n^{o(n)}\poly q$ 
and thus contradict Lemma~\ref{lem:hies2-hard} under \#ETH. $\qed$ 

\subsection{Proof of Theorem~\ref{thm:main-lb-fixed}}
Fix an arbitrary prime power $q_0$. 
We will rely on Lemma~\ref{lem:hsies2-hard} and Theorem~\ref{thm:critical}. 
Let $\varphi$ be a \#$(q,3,2(n+q),O(q^2\polylog q))$-CSP instance with 
the structure of a homogeneous sum-inequation system over $\F_q$ 
with $q=q_0^d=n^{1+o(1)}$. 

Construct a $k\times m$ matrix $G\in\F_{q_0}^{k\times m}$ with $k=2(n+q)$ so
that each column of $G$ corresponds to a unique sum-inequation of $\varphi$; 
in particular, every column of $G$ has at most three nonzero entries.
Recalling Lemma~\ref{lem:counting-extension-full-support} and 
the construction in Sect.~\ref{sect:techniques}, 
extend $G$ elementwise from $\F_{q_0}$ to $\F_q=\F_{q_0^d}$ to obtain 
$\bar G\in\F_{q}^{k\times m}$.
For all $\bar x\in\F_q^k$ we have that $\bar x\bar G$ has full support 
if and only if $\bar x\in\SAT(\varphi)$. Theorem~\ref{thm:critical} 
and Lemma~\ref{lem:counting-extension-full-support}
thus imply that $(-1)^{\rho(G)}T_G(1-q_0^d,0)=|\SAT(\varphi)|$.
Since $m=O(q^2\polylog n)$, we have $m=k^{O(1)}$. 
An algorithm that computes the Tutte polynomial $T_G$ in time $k^{o(k)}$
would thus enable us to compute $|\SAT(\varphi)|$ in time $n^{o(n)}\poly q$ 
and thus contradict Lemma~\ref{lem:hsies2-hard} under \#ETH. $\qed$

\section{An upper bound for the general case}
\label{sect:ub}

This section proves our first upper-bound result, 
Theorem~\ref{thm:main-ub}.
Let $\F$ be a field and let $M\in\F^{k\times m}$ be a $k\times m$ matrix 
with columns indexed by a set $E$ with $|E|=m$ given as input. 
Our task is to compute the Tutte polynomial $T_M(x,y)$ in coefficient form. 

\subsection{Least generators and prefix-dependent partitioning}

We start with preliminaries towards Theorem~\ref{thm:main-ub}.
Let us assume that the set $E$ is totally ordered. 
For two distinct subsets $A,B\subseteq E$, we say that $A$ is 
{\em size-lexicographically lesser than} $B$ and write $A<B$ if
either $|A|<|B|$ or both $|A|=|B|$ and the minimum element of 
$(A\setminus B)\cup (B\setminus A)$ belongs to $A$.

For a set $S\subseteq E$, let us write $L(S)$ for the size-lexicographically 
least subset of $S$ such that $\rho(L(S))=\rho(S)$. We say that $L(S)$ is the 
{\em least generator set} for $S$; indeed, $M[L(S)]$ generates the column 
space of $M[S]$. Furthermore, we observe that $|L(S)|=\rho(L(S))$; indeed,
otherwise we would have $|L(S)|>\rho(L(S))=\rho(S)$, which would mean that
there would exist an $e\in L(S)$ with 
$\rho(L(S)\setminus\{e\})\geq\rho(L(S))=\rho(S)$, in which case
$L(S)\setminus\{e\}$ would contradict the size-lexicographic leastness 
of $L(S)$. In particular, $L(S)$ is an independent set. 

For an independent set $I\subseteq E$, let us say that an element $f\in E$
is {\em $I$-prefix-dependent} if $M[f]$ is in the column span of 
$M[\{e\in I:e<f\}]$. Let us write $P(I)$ for the set of all 
$I$-prefix-dependent elements of $E$. We observe that given $I$ as input, 
$P(I)$ can be computed in $\poly(k,m)$ operations in $\F$.

\begin{Lem}[Prefix-dependent partitioning]
\label{lem:prefix-dep}
For all $S\subseteq E$ it holds that 
\[
L(S)\subseteq S\subseteq L(S)\cup P(L(S))\,, 
\]
where the union is disjoint.
\end{Lem}
\begin{Proof}
Let us first observe that the union must be disjoint; indeed, every element
of $P(L(S))$ depends on one or more of elements of $L(S)$, and 
$L(S)$ is independent. The inclusion $L(S)\subseteq S$ is immediate by the
definition of $L(S)$. Next, observe that $S\subseteq L(S)\cup P(L(S))$
holds trivially when $S$ is the empty set, so let us assume $S$ is nonempty.
Consider an arbitrary $f\in S$. If $f\in L(S)$, we are done. 
So suppose that $f\notin L(S)$. Since $M[L(S)]$ generates the column space
of $M[S]$, we have that $M[f]$ depends on $M[K]$ for some 
$f\notin K\subseteq L(S)$. Take the size-lexicographically least such $K$. 
If $e<f$ holds for all $e\in K$, we have $f\in P(L(S))$ and we are done.
So suppose that there is an $e\in K$ with $f<e$. By size-lexicographic leastness
of $K$, $M[f]$ is not in the span of $M[K\setminus\{e\}]$; that is,
$M[K\cup\{f\}\setminus\{e\}]$ is independent, and thus must generate the
same space as $M[K]$. Since $K\subseteq L(S)$ and $f\in S\setminus L(S)$, 
it follows that $L(S)\cup\{f\}\setminus\{e\}$ contradicts the 
size-lexicographic leastness of $L(S)$, and the lemma follows.
\end{Proof}

\subsection{Computing the Tutte polynomial via least generator sets}

This section completes our proof of Theorem~\ref{thm:main-ub}.
The key idea in our algorithm is now to implement the contribution of each 
set $S\subseteq E$ to the Tutte polynomial through 
the least generator set $L(S)$ and the associated 
{\em prefix-dependent residual}
$R=S\setminus L(S)\subseteq P(L(S))$ enabled by Lemma~\ref{lem:prefix-dep}. 
Indeed, $L(S)$ is independent, which enables us to work over only the 
independent sets $I$ of $M$, each of which has size at most $k$. 
More precisely, let us write $\binom{E}{\ell}$ for the set of all 
$\ell$-element subsets of $E$. From the definition \eqref{eq:tutte} of 
the Tutte polynomial and Lemma~\ref{lem:prefix-dep}, we immediately have
\begin{equation}
\label{eq:tutte-via-residual}
\begin{split}
T_M(x,y)
&=\sum_{S\subseteq E}(x-1)^{k-\rho(S)}(y-1)^{|S|-\rho(S)}\\
&=\sum_{\ell=0}^k
    \sum_{\substack{I\in\binom{E}{\ell}\\\rho(I)=\ell}}
      (x-1)^{k-\ell}\sum_{R\subseteq P(I)}(y-1)^{\ell+|R|-\ell}\\
&=\sum_{\ell=0}^k
    \sum_{\substack{I\in\binom{E}{\ell}\\\rho(I)=\ell}}
      (x-1)^{k-\ell}y^{|P(I)|}\,,
\end{split}
\end{equation}
where the last equality follows from the Binomial Theorem.
It follows from \eqref{eq:tutte-via-residual} that we can compute
$T_M(x,y)$ by iterating over the subsets of $E$ of size at most $k$,
using at most $\poly(m,k)$ arithmetic operations in $\F$ in each iteration.
When $m=k^{O(1)}$ and $\F$ is a finite field, Theorem~\ref{thm:main-ub} follows
since there are at most $km^k=k^{O(k)}$ such subsets and each arithmetic 
operation in $\F_q$ can be implemented in time $\polylog q$ 
(cf.~\cite{vonzurGathenG2013}). $\qed$

\section{An upper bound for weight at most two}
\label{sect:ub-wt2}

This section proves Theorem~\ref{thm:main-ub-wt2}.
Let us assume that the field $\F$ has $q$ elements.
Furthermore, let us assume that every column of the given input 
$M\in\F^{k\times m}$ has at most two nonzero elements;
without loss of generality we may assume that $M$ has no all-zero columns. 
Let us index the set of rows of $M$ by a set $V$ with $|V|=k$
and the set of columns by a set $E$ with $|E|=m$.

\subsection{Multigraphs and the two possible ranks in the connected case}

Our strategy is to derive counting recurrences for the coefficients 
of the Tutte polynomial using a multigraph representation of $M$. 
Indeed, the pair $(V,E)$ together with $M$ naturally defines a multigraph $G$ 
with vertex set $V$ and edge set $E$ such that each edge $e\in E$ is either (i) 
a loop at vertex $v\in V$ if the only nonzero entry at column $e$ of $M$ is
$m_{ve}$ or (ii) an edge joining two distinct vertices $v,w\in V$ if the 
nonzero entries at column $e$ of $M$ are $m_{ve}$ and $m_{we}$. 

To compute the Tutte polynomial $T_M(x,y)$ in coefficient form, it is
immediate from \eqref{eq:tutte} that it 
suffices to have available the following coefficients. For $r=0,1,\ldots,k$ 
and $s=0,1,\ldots,m$, define the coefficient
\begin{equation}
\label{eq:tau}
\tau_{r,s}=\bigl|\bigr\{S\subseteq E:\rho(S)=r\,,\ |S|=s\bigr\}\bigr|\,.
\end{equation}
Our approach on computing the coefficients $\tau_{r,s}$ will be based on 
the structure of the multigraphs $(V,S)$ with vertex set $V$ and edge set 
$S\subseteq E$. Towards this end, for a nonempty subset $U\subseteq V$,
it will be convenient to write $S[U]$ for the set of all edges $e\in S$
incident only to vertices in $U$. Suppose now that $U\subseteq V$ is the
vertex set of a connected component of $(V,S)$. Then, it is immediate that
$S[U]$ is the edge set of this connected component, and well-known that the 
rank of the submatrix $M[S[U]]$ is either $|U|-1$ or $|U|$. To see the
latter, first observe that $M[S[U]]$ has at most $|U|$ nonzero rows, so the
rank is at most $|U|$. Next, observe that the multigraph $(U,S[U])$ is 
connected, so it has a spanning tree of $|U|-1$ edges; select a root vertex 
and orient the edges of the spanning 
tree so that the head vertex of each arc has greater distance to the root than 
the tail vertex; perform a topological sorting of the resulting directed 
acyclic graph, and observe that Gaussian elimination applied to $M[S[U]]$ in 
this topological order of rows---for each arc, use the head to zero out the 
tail---leaves a reduced echelon form with at least $|U|-1$ 
independent columns of weight one. Thus, the rank of $M[S[U]]$ is 
at least $|U|-1$, and hence either $|U|-1$ or $|U|$. 
We proceed to derive counting recurrences that distinguish between these 
two possible ranks and perform dynamic programming over vertex sets of 
connected components. 

\subsection{Preliminaries: Counting subgraphs by the number of components and edges}

For ease of exposition, 
let us first derive a counting recurrence for spanning subgraphs that does not 
distiguish between the ranks but explicitly tracks the number of connected 
components and the number of edges. We emphasize that this recurrence is 
known and due to Bj\"orklund~{\em et al.}~\cite{BjorklundHKK2008}.
For a multigraph with vertex set $U\subseteq V$ and edge set 
$S\subseteq E[U]$, let us write $c(U,S)$ for the number of connected 
components in $(U,S)$.

For nonempty $U\subseteq V$, $d=1,2,\ldots,k$, and 
$s=0,1,\ldots,m$, define
\begin{equation}
\label{eq:al}
\alpha_{d,s}(U)=
\bigl|\bigl\{S\subseteq E[U]:c(U,S)=d\,,\ |S|=s\bigr\}\bigr|\,.
\end{equation}
That is, $\alpha_{d,s}(U)$ counts the number of $U$-spanning subgraphs 
of $G$ with exactly $d$ connected components and $s$ edges. 
We observe that $\alpha_{d,s}(U)=0$ unless both $d\leq |U|$ and $s\leq |E[U]|$.
From \eqref{eq:al} it is thus immediate that
\[
\alpha_{1,s}(U)+\alpha_{2,s}(U)+\ldots+\alpha_{|U|,s}(U)=\binom{|E[U]|}{s}\,.
\]
Put othewise, the connected case $d=1$ can be solved via the
disconnected cases $d=2,3,\ldots,|U|$ by
\begin{equation}
\label{eq:al-sub}
\alpha_{1,s}(U)=\binom{|E[U]|}{s}-\alpha_{2,s}(U)-\ldots-\alpha_{|U|,s}(U)\,.
\end{equation}
The disconnected cases $d=2,3,\ldots,|U|$ can in turn can be solved via the 
cases $W\subsetneq U$.
Indeed, let us observe that 
\begin{equation}
\label{eq:al-conv}
\alpha_{d,s}(U)=\frac{1}{d}\sum_{t=0}^s\sum_{\emptyset\neq W\subsetneq U} \alpha_{1,t}(W)\alpha_{d-1,s-t}(U\setminus W)\,.
\end{equation}
To justify \eqref{eq:al-conv}, 
observe that an arbitrary $d$-component subgraph with $s$ 
edges and vertex set $U$ has exactly $d$ choices for the vertex set $W$
of a connected component; this connected component with vertex set $W$ has 
$t$ edges for exactly one choice of $t=0,1,\ldots,s$, leaving $s-t$ 
edges and the vertex set $U\setminus W$ for the other 
$d-1$ components; moreover, the choices for the subgraph on $W$ and the 
subgraph on $U\setminus W$ are independent of each other. 

\subsection{Partitioning the connected case by rank via hyperplane sieving}

We now extend the recurrence \eqref{eq:al} to distinguish between the
ranks $|U|-1$ and $|U|$ in the connected case $\alpha_{1,s}(U)$. 
That is, we will partition the $\alpha_{1,s}(U)$ sets $S\subseteq E[U]$ 
into two classes: 
\begin{itemize}
\item[(i)]  
the sets $S$ for which $M[S]$ has full rank $|U|$, and 
\item[(ii)] 
the sets $S$ for which $M[S]$ has rank $|U|-1$; that is, 
a rank-deficiency of one from 
full rank. 
\end{itemize}
From the Rank--Nullity Theorem, we know that the null space of a matrix 
has dimension one if the matrix has a rank deficiency of one from full rank, 
whereas the null space is trivial if the matrix has full rank. 
Hence, we observe that case 
(ii) occurs if and only if every column of the matrix $M[S]$ belongs to 
a hyperplane $\sum_{v\in V}h(v)x_v=0$ 
defined by a not-identically-zero function $h:V\rightarrow\F$ that is 
unique up to multiplication by a nonzero scalar. Moreover, since the 
multigraph $(U,S)$ is connected, we observe that the support 
$H=\{v\in V:h(v)\neq 0\}$ must contain the set $U$; 
indeed, otherwise by connectedness there exists an edge in $e\in S$ that joins 
a vertex $w_0\in H$ with a vertex $w_1\in U\setminus H$, which is a 
contradiction since then
$\sum_{v\in V}h(v)m_{ve}=h(w_0)m_{w_0e}+h(w_1)m_{w_1e}=h(w_0)m_{w_0e}\neq 0$
and thus the column $M[e]$ does not lie in the hyperplane defined by $h$.
Accordingly, we may assume that in case (ii) the hyperplane is defined by 
a function $h:U\rightarrow\F\setminus\{0\}$ that is unique up to multiplication
by a nonzero scalar.

Our strategy is now to count the case (ii) by sieving over all possible 
functions $h:U\rightarrow\F\setminus\{0\}$, restricting \eqref{eq:al} 
accordingly for each choice of $h$, and finally to compensate for the overcount
by dividing with the number $q-1$ of nonzero scalars in $\F$. 
Towards this end, for nonempty $U\subseteq V$, 
$h:U\rightarrow\F\setminus\{0\}$, and $S\subseteq E[U]$, define the 
$h$-{\em restriction} of $S$ by
\begin{equation}
\label{eq:}
S^h=\biggl\{e\in S:\sum_{v\in U}h(v)m_{ve}=0\biggr\}\,.
\end{equation}
We obtain the $h$-restricted version of
\eqref{eq:al} by defining, for $d=1,2,\ldots,k$ and $s=0,1,\ldots,m$, 
\begin{equation}
\label{eq:al-h}
\alpha_{d,s}^h(U)=
\bigl|\bigl\{S\subseteq E[U]^h:c(U,S)=d\,,\ |S|=s\bigr\}\bigr|\,.
\end{equation}
We observe that $\alpha^h_{d,s}(U)=0$ unless both $d\leq |U|$ and 
$s\leq |E[U]^h|$.
From \eqref{eq:al-h} it is immediate that
\[
\alpha_{1,s}^h(U)+\alpha_{2,s}^h(U)+\ldots+\alpha_{|U|,s}^h(U)=\binom{|E[U]^h|}{s}\,.
\]
Similarly to \eqref{eq:al-sub}, the connected case $d=1$ can be solved via the
disconnected cases $d=2,3,\ldots,|U|$ by
\begin{equation}
\label{eq:al-h-sub}
\alpha_{1,s}^h(U)=
\binom{|E[U]^h|}{s}-\alpha_{2,s}^h(U)-\ldots-\alpha_{|U|,s}^h(U)\,.
\end{equation}
Similarly to \eqref{eq:al-conv}, 
the disconnected cases $d=2,3,\ldots,|U|$ can in turn can be solved via the 
cases $W\subsetneq U$. For nonempty $W\subseteq U$ and 
$h:U\rightarrow\F\setminus\{0\}$, let us write $h_W$ for the restriction 
of $h$ to $W$. We have
\begin{equation}
\label{eq:al-h-conv}
\alpha_{d,s}^h(U)=\frac{1}{d}\sum_{t=0}^s\sum_{\emptyset\neq W\subsetneq U} \alpha_{1,t}^{h_W}(W)\alpha_{d-1,s-t}^{h_{U\setminus W}}(U\setminus W)\,.
\end{equation}
We are now ready to split into the cases (i) and (ii). Since 
$h:U\rightarrow\F\setminus\{0\}$ is unique up to multiplication by a nonzero
scalar, 
for nonempty $U\subseteq V$ and $s=0,1,\ldots,m$, we have that
case (ii) is counted by
\begin{equation}
\label{eq:beii}
\beta^{\text{(ii)}}_s(U)=\frac{1}{q-1}\sum_{h:U\rightarrow\F\setminus\{0\}}\alpha_{1,s}^h(U)\,.
\end{equation}
Case (i) is thus counted by
\begin{equation}
\label{eq:bei}
\beta^{\text{(i)}}_s(U)=\alpha_{1,s}(U)-\beta^{\text{(ii)}}_s(U)\,.
\end{equation}

\subsection{Counting by rank and number of edges via connected components}

Let us next use the coefficients $\beta^{\text{(i)}}_s(U)$ and $\beta^{\text{(ii)}}_s(U)$ to derive
a recurrence for the coefficients $\tau_{r,s}$ in \eqref{eq:tau}. 
Let us write $c^{\text{(i)}}(U,S)$ 
and $c^{\text{(ii)}}(U,S)$ for the number of components of type (i) and
(ii) in $(U,S)$, respectively. 
For nonempty $U\subseteq V$, 
$d^{\text{(i)}}=0,1,\ldots,k$,
$d^{\text{(ii)}}=0,1,\ldots,k$, and
$s=0,1,\ldots,m$, define
\begin{equation}
\label{eq:sigma}
\sigma_{d^{\text{(i)}},d^{\text{(ii)}},s}(U)=
\bigl|\bigl\{S\subseteq E[U]:c^{\text{(i)}}(U,S)=d^{\text{(i)}}\,,\ c^{\text{(ii)}}(U,S)=d^{\text{(ii)}}\,,\ |S|=s\bigr\}\bigr|\,.
\end{equation}
We observe that $\sigma_{d^{\text{(i)}},d^{\text{(ii)}},s}(U)=0$
unless both $1\leq d^{\text{(i)}}+d^{\text{(ii)}}\leq |U|$ and $s\leq |E[U]|$.
Since each component of type (ii) contributes a rank-deficiency of one, 
from \eqref{eq:sigma} and \eqref{eq:tau} we observe that 
\begin{equation}
\label{eq:tau-sum}
\tau_{r,s}=\sum_{d^{\text{(i)}}=0}^k\sigma_{d^{\text{(i)}},k-r,s}(V)\,, 
\end{equation}
so it suffices to have a recurrence for
computing the coefficients \eqref{eq:sigma}. 

Towards this end, let $U\subseteq V$ be nonempty.
For $d^{\text{(i)}}=0,1,\ldots,|U|$, $d^{\text{(ii)}}=0,1,\ldots,|U|$,
and $s=0,1,\ldots,|E[U]|$, we have
\begin{equation}
\label{eq:sigma-conv}
\small
\sigma_{d^{\text{(i)}},d^{\text{(ii)}},s}(U)
=
\begin{cases}
\beta^{\text{(i)}}_s(U) & \text{if $d^{\text{(i)}}=1$ and $d^{\text{(ii)}}=0$};\\[1mm]
\beta^{\text{(ii)}}_s(U) & \text{if $d^{\text{(i)}}=0$ and $d^{\text{(ii)}}=1$};\\[1mm]
\frac{1}{d^{\text{(i)}}}
\sum_{t=0}^s
\sum_{\emptyset\neq W\subsetneq U}
   \beta^{\text{(i)}}_t(W)\sigma_{d^{\text{(i)}}-1,d^{\text{(ii)}},s-t}(U\setminus W)
&\text{if $d^{\text{(i)}}\geq 2$ or $d^{\text{(i)}},d^{\text{(ii)}}\geq 1$};\\[1mm]
\frac{1}{d^{\text{(ii)}}}
\sum_{t=0}^s
\sum_{\emptyset\neq W\subsetneq U}
   \beta^{\text{(ii)}}_t(W)\sigma_{d^{\text{(i)}},d^{\text{(ii)}}-1,s-t}(U\setminus W)
&\text{if $d^{\text{(ii)}}\geq 2$ or $d^{\text{(i)}},d^{\text{(ii)}}\geq 1$}.
\end{cases}
\end{equation}
Indeed, the multigraph $(U,S)$ for an arbitrary $S\subseteq E[U]$ 
with $|S|=s$ splits uniquely into 
$d^{\text{(i)}}$ and $d^{\text{(ii)}}$ connected components of types
(i) and (ii), respectively. When $d^{\text{(i)}}+d^{\text{(ii)}}=1$,
this connected component is unique and enumerated either by 
$\beta^{\text{(i)}}_s(U)$ or by $\beta^{\text{(ii)}}_s(U)$. 
When $d^{\text{(i)}}\geq 2$ or $d^{\text{(i)}},d^{\text{(ii)}}\geq 1$,
there are exactly $d^{\text{(i)}}$ choices for the vertex set $W$ 
of a connected component of type~(i); this connected component with 
vertex set $W$ has $t$ edges for exactly one choice of $t=0,1,\ldots,s$,
leaving $s-t$ edges and the vertex set $U\setminus W$ for the other
$d^{\text{(i)}}-1$ components of type (i) as well as the 
$d^{\text{(ii)}}$ components of type (ii). 
When $d^{\text{(ii)}}\geq 2$ or $d^{\text{(i)}},d^{\text{(ii)}}\geq 1$,
there are exactly $d^{\text{(ii)}}$ choices for the vertex set $W$ 
of a connected component of type~(ii); this connected component with 
vertex set $W$ has $t$ edges for exactly one choice of $t=0,1,\ldots,s$,
leaving $s-t$ edges and the vertex set $U\setminus W$ for the other
$d^{\text{(ii)}}-1$ components of type (ii) as well as the 
$d^{\text{(i)}}$ components of type (i).

\subsection{Fast evaluation of the recurrences}

This section completes the proof of Theorem~\ref{thm:main-ub-wt2}.
By assumption, we have $m=k^{O(1)}$.
Recall also that we write $q$ for the number of elements in the finite
field $\F$. 
It remains to show that we can compute the coefficients 
$\sigma_{r,s}(V)=\tau_{r,s}$ for $r=0,1,\ldots,k$ and $s=0,1,\ldots,m$
in time and space $q^kk^{O(1)}$. 

Following Bj\"orklund~{\em et al.}~\cite{BjorklundHKK2008}, we recall
that in time $2^kk^{O(1)}$ we can compute the coefficients $\alpha_{d,s}(U)$ 
in~\eqref{eq:al} for all nonempty $U\subseteq V$, $d=1,2,\ldots,k$, and 
$s=0,1,\ldots,m$. The computation proceeds one {\em level} 
$\ell=1,2,\ldots,k$ at a time, where at level $\ell$ we solve for all the 
coefficients $\alpha_{d,s}(U)$ with $|U|=\ell$. 
The base case $\ell=1$ with $\alpha_{1,s}(U)=\binom{|E[U]|}{s}$ is
immediate. For $\ell\geq 2$, we assume the values at all the previous levels
$1,2,\ldots,\ell-1$ have already been computed and proceed as follows. 
First, for each $s=0,1,\ldots,m$, we apply fast subset 
convolution~\cite{BjorklundHKK2007} on \eqref{eq:al-conv} 
for each $t=0,1,\ldots,s$ using the already 
computed values to obtain the disconnected cases $d=2,3,\ldots,\ell$ for 
all $U\subseteq V$ with $|U|=\ell$. 
Then, for each $U\subseteq V$ with $|U|=\ell$
and $s=0,1,\ldots,m$, we solve for the connected case $d=1$ 
using \eqref{eq:al-sub}. Since $m=k^{O(1)}$, this takes 
$2^kk^{O(1)}$ time in total. Since there are $k$ levels, 
we have that all the coefficients $\alpha_{d,s}(U)$ can be computed in
$2^kk^{O(1)}\leq q^kk^{O(1)}$ total time. 

Next, let us study the $h$-restricted coefficients $\alpha_{d,s}^h(U)$ in
\eqref{eq:al-h}. First, we observe from the Binomial Theorem that there
are exactly $q^k=(q-1+1)^k=\sum_{\ell=0}^k\binom{k}{\ell}(q-1)^\ell 1^{k-\ell}$
functions $h:U\rightarrow\F\setminus\{0\}$ with $U\subseteq V$.
Accordingly and with foresight, let us work with the following one-to-one 
correspondence. For nonempty $U\subseteq V$, 
identify a function $h:U\rightarrow\F\setminus\{0\}$
with the nonzero vector $\bar h\in\F^V\cong\F^k$ defined for all 
$v\in V$ by
\begin{equation}
\label{eq:h-bar}
\bar h_v=\begin{cases}
h(v) & \text{if $v\in U$};\\
0    & \text{otherwise}.
\end{cases}
\end{equation}
In particular, the domain $U$ of $h$ is exactly the support of $\bar h$. 
Thus, we can write $\bar\alpha_{d,s}(\bar h)=\alpha_{d,s}^h(U)$ and accordingly 
view $\bar\alpha_{d,s}:\F^k\rightarrow\Z$ as a function that 
assigns an integer value $\bar\alpha_{d,s}(\bar h)$ to each $\bar h\in\F^k$,
with the tacit convention that the all-zero vector is assigned to zero.

To compute the values of $\bar\alpha_{d,s}$ on each nonzero vector 
$\bar h\in\F^k$,
let us again proceed one level $\ell=1,2,\ldots,k$ at a time, where at level 
$\ell$ we solve for all the values $\bar\alpha_{d,s}(\bar h)$ where $\bar h$
has exactly $\ell$ nonzero entries. The base case $\ell=1$ with 
$\bar\alpha_{1,s}(\bar h)=\binom{|E[U]^h|}{s}$ is immediate. 
For $\ell\geq 2$, we assume the values at all the previous levels
$1,2,\ldots,\ell-1$ have already been computed and proceed as follows.
Introduce an arbitrary total order $\leq$ into $\F$ with the property 
that $0\in\F$ is the minimum element in this order. Partially order $\F^k$
by taking the direct product of $k$ copies of $(\F,\leq)$. This partial
order is a lattice with $q^k$ elements, $(q-1)k$ of which are join-irreducible. 
In particular, this enables us to compute the join-product of two
functions $\bar\phi,\bar\psi:\F^k\rightarrow\Z$ in $q^{k+1}k^{O(1)}$ 
arithmetic operations using 
fast M\"obius inversion~\cite{BjorklundHKKNP2016}. More precisely, for 
two vectors $\bar f,\bar g\in\F^k$, define the {\em join} 
$\bar f\vee \bar g\in\F^k$ to be the 
element-wise $\leq$-maximum of $\bar f$ and $\bar g$. The {\em join-product} 
$\bar\phi\vee\bar\psi:\F^k\rightarrow\Z$ is defined for all $h\in\F^k$ by
\begin{equation}
\label{eq:join-prod}
(\bar\phi\vee\bar\psi)(\bar h)=\sum_{\bar f,\bar g\in\F^k:\bar f\vee \bar g=\bar h}\bar\phi(\bar f)\bar\psi(\bar g)\,. 
\end{equation}
For a function $\bar\phi:\F^k\rightarrow\Z$ and $\ell=0,1,\ldots,k$, define 
the {\em weight} $\ell$ {\em part} of $\bar\phi$ to be the function 
$[\bar\phi]_\ell:\F^k\rightarrow\Z$ defined for all $\bar h\in\F^k$
by
\begin{equation}
\label{eq:wt-ell-part}
[\bar\phi]_\ell(\bar h)=
\begin{cases}
\bar\phi(\bar h) & \text{if $\bar h$ has exactly $\ell$ nonzero entries};\\
0 & \text{otherwise}.
\end{cases}
\end{equation}
From the correspondence \eqref{eq:h-bar}, the definitions
\eqref{eq:join-prod} and \eqref{eq:wt-ell-part}, as well as the 
recurrence~\eqref{eq:al-h-conv}, we thus observe that
for $d=2,3,\ldots,k$ and $s=0,1,\ldots,m$ we have 
\begin{equation}
\label{eq:bar-al-h-conv}
\bigl[\bar\alpha_{d,s}\bigr]_\ell
=
\frac{1}{d}\sum_{j=1}^{\ell-1}\sum_{t=0}^s
\bigl[\bigl[\bar\alpha_{1,t}\bigr]_j\vee
\bigl[\bar\alpha_{d-1,s-t}\bigr]_{\ell-j}\bigr]_\ell\,.
\end{equation}
In particular, using fast join products~\cite{BjorklundHKKNP2016} on
\eqref{eq:bar-al-h-conv}, we can solve for the disconnected cases
$d=2,3,\ldots,k$ for all $s=0,1,\ldots,m$, $h:U\rightarrow\F\setminus\{0\}$ 
and $U\subseteq V$ with $|U|=\ell$ in total time $q^{k+1}k^{O(1)}$. We can thus 
solve the connected case $d=1$ via \eqref{eq:al-h-sub} to achieve total time 
$q^{k+1}k^{O(1)}$ at level $\ell$. Since there are $k$ levels, we achieve total
time $q^{k+1}k^{O(1)}$ to compute all the values $\alpha_{d,s}^h(U)$ and hence
all the values $\beta_s^{\text{(i)}}(U)$ and $\beta_s^{\text{(ii)}}(U)$ 
via \eqref{eq:bei} and \eqref{eq:beii}. 
Finally, we solve for the coefficients $\tau_{r,s}$ in \eqref{eq:tau}
using \eqref{eq:tau-sum} and the recurrence \eqref{eq:sigma-conv}
for the coefficients $\sigma_{d^{\text{(i)}},d^{\text{(ii)}},s}$ in
\eqref{eq:sigma}. In particular, the recurrence \eqref{eq:sigma-conv} can
be evaluated in time $2^kk^{O(1)}$ using fast subset 
convolution~\cite{BjorklundHKK2008}. Thus, the entire algorithm runs in
time $q^{k+1}k^{O(1)}$.
This completes the proof of Theorem~\ref{thm:main-ub-wt2}.
$\qed$

%%%%%%%%%%%%%%%%%%%%%%%%%%%%%%%%%%%%%%%%%%%%%%%%%%%%%%%%%%% Acknowledgments %%%

\section*{Acknowledgments}

We thank the anonymous reviewers of an earlier version of this manuscript for their useful remarks and questions that led to Theorem~\ref{thm:main-ub-wt2}.

%%%%%%%%%%%%%%%%%%%%%%%%%%%%%%%%%%%%%%%%%%%%%%%%%%%%%%%%%%%%%%%% References %%%

\bibliographystyle{abbrv}
\bibliography{paper.bib}

%%%%%%%%%%%%%%%%%%%%%%%%%%%%%%%%%%%%%%%%%%%%%%%%%%%%%%%%%%%%% Document ends %%%

\end{document}